\documentstyle[aps,prd,eqsecnum,multicol,fancyheadings]{revtex}

\def\bbbc{{\mathchoice {\setbox0=\hbox{$\displaystyle\rm C$}\hbox{\hbox
to0pt{\kern0.4\wd0\vrule height0.9\ht0\hss}\box0}}
{\setbox0=\hbox{$\textstyle\rm C$}\hbox{\hbox
to0pt{\kern0.4\wd0\vrule height0.9\ht0\hss}\box0}}
{\setbox0=\hbox{$\scriptstyle\rm C$}\hbox{\hbox
to0pt{\kern0.4\wd0\vrule height0.9\ht0\hss}\box0}}
{\setbox0=\hbox{$\scriptscriptstyle\rm C$}\hbox{\hbox
to0pt{\kern0.4\wd0\vrule height0.9\ht0\hss}\box0}}}}
\input psfig.tex
\begin{document}
\draft
\title {Bound systems in an expanding universe}
\author{George A. Baker, Jr.}
\address{
Theoretical Division, Los Alamos National Laboratory\\
University of California, Los Alamos, N. M. 87544 USA }
\date{submitted June 12, 2000}
\maketitle
\begin {abstract}
The Schwarzschild solution insertion in an expanding universe model,
the so called ``Swiss cheese model,'' is shown to possess a very
unphysical property. Specifically, in this model some of the trajectories
are discontinuous functions of their initial conditions.
An alternate metric is proposed as a remedy. It goes
smoothly between the Schwarzschild exterior solution and the
Friedmann-Lema\^\i tre, expanding universe metric. It is
further shown that the effects of the expansion on planetary motions
in the solar system are too small to be currently observed for this alternate
metric.
\end {abstract}
\pacs{95.30Sf, 04.20-q, 04.40-b, 04.20Jb}

\vspace*{-0.4cm}
\begin{multicols}{2}

\columnseprule 0pt

\narrowtext
\section{INTRODUCTION AND SUMMARY}
Currently there are two general relativistic descriptions of spacetime
in popular use.  For planetary systems and other gravitationally
bound structures which are small on  the scale of the universe,
there is a static description of the behavior of spacetime. On the other
hand, for large-scale behavior, there is a time dependent description
which is appropriate as a description of phenomena such as the observed
red-shift of distant galaxies.

The classic question is, ``How can these two disparate
descriptions of spacetime possibly be reconciled with each other?'' The
current standard answer is that these two are meshed together\cite{Peb} on a
spherical surface surrounding a mass concentration which grows with
time in the static metric, but stays at a fixed coordinate radius in the
non-static metric.  This is the ``Schwarzschild solution in a cosmological
model'' picture, or it is also called the ``Swiss cheese model.'' This
later name refers to the fact the in this picture, the background material
is removed inside the spherical boundary.  The mass removed depends on
the mass of the central concentration and the curvature of space.
The replacement of the mass interior to the sphere by
a concentration of mass at the center is based on Birkhoff's theorem\cite{Br}
which says that in a homogeneous, zero pressure cosmological model, as
long as the material inside a sphere is spherically symmetric, we can replace
it with a compact mass at the center with no change on its exterior effects.
To
make this matching work, the pressure of the exterior solution (Friedmann-%
Lema\^\i tre metric class of the general Robertson-Walker line elements)
must vanish, as does the pressure of the interior (the exterior Schwarzschild
metric) solution.  This condition places a restriction on the form of
the universal expansion factor of the overall universe.  Although the
matching conditions can be met, as we shall see in the third section, an
additional problem arises when the dynamics are considered.

In the second section, for the convenience of the reader, I gather together
the necessary classical equations from general relativity for the study at
hand.  One non-classical item in this section is, instead of the usual
$3+1$ spacetime split into 3 space and 1 time dimensions, I split
spacetime into one radial coordinate, and time plus the two angles of
spherical coordinates.  The change allows the direct computation of the
relevant extrinsic curvatures.

In the third section, I compute the stress-energy tensors and the extrinsic
curvatures for both the Schwarzschild and the Friedman-Lema\^\i tre metrics
used in the ``Swiss cheese model.''  By a reparameterization of the
Schwarzschild metric, both the intrinsic and the extrinsic curvature
can be made to be continuous.
I remark that the stress-energy tensor, for certain parameter choices,
displays no pressure discontinuity but only a cosmic fluid density
discontinuity. That discontinuity is in line with the ``Swiss cheese model''
idea that there are holes in the cosmic fluid.

In the fourth section, I show that it can happen, for trajectories which
approach the metric interface an near grazing angles, that the subsequent
trajectories are discontinuous functions of their initial conditions. Those
which enter the inner or Schwarzschild metric region can be bound in
a {\it finite sized}, closed orbit, while those which do not, travel on
a parabolic trajectory to infinity.  To emphasis, this case is not the
same as in Newtonian orbit theory where ellipses of progressively larger
size blend into parabolas, but here the ellipse is just finite in size!

In the fifth section, I introduce an alternative metric.  This metric is
basically an adaptation of the Schwarzschild metric in curved space.
I compute the stress-energy tensor. It shows an isotropic pressure, and
no mass-density flux.  It is only second order in magnitude in both
the Hubble constant and the inverse radius of curvature of the universe.
The same statement is true of the Friedmann-Lema\^\i tre stress energy
tensor.  In addition I have computed the extrinsic curvature.
Both the stress-energy tensor and the extrinsic curvature are continuous,
outside the Schwarzschild radius of course, as they come from infinitely
differentiable expressions.

In the sixth section, I compute the equations of motion for a freely
moving test particle in the alternative metric.  I then transform them
to a coordinate system at rest at the center of the mass concentration.
The flat space, slowly moving particle, weak gravitational field limit
of these equations of motion are also given.  The only correction to
Newton's equations of motion in this limit is a term proportional to
the square of Hubble's constant, $H_0$.

In the final section, I gives some examples of the dynamics found
using my alternative metric.

\section{METRICS}

It is useful to review the properties of a few metrics.  The line elements
will all be of the general form,
\begin{eqnarray}
ds^2&=& -e^\mu[(dx^1)^2+(dx^2)^2+(dx^3)^2]+e^\nu (dx^4)^2 \nonumber \\
&=&g_{ij}dx^idx^j, \label{2.1}
\end{eqnarray}
where the Einstein summation convention is used, and where
\begin{equation}
r^2=(x^1)^2+(x^2)^2+(x^3)^2,\quad \mu =\mu(r,t),\quad \nu =\nu (r,t),
\label{2.2}\end{equation}
appropriate to the non-static, spherically symmetric case. Eq.~\ref{2.1}
defines the metric tensor $g_{ij}$.
It will be of interest to know what Einstein field equation is satisfied
for each of these metrics. The equation is,
\begin{equation}
R_i^j-{\textstyle {1\over 2}}Rg_i^j +\Lambda g_i^j=-8\pi T_i^j, \label{2.3}
\end{equation}
where  $T_{ij}$ is the
stress-energy tensor, and $R_{ij}$ is the Ricci tensor, which is a
contraction of Riemann's four index curvature tensor.  The Ricci
tensor can be expressed in terms of the three index Christoffel
symbols $\Gamma $ as,
\begin{equation}
R_{km}=(\Gamma ^i_{km})_{,i}-(\Gamma ^i_{ki})_{,m}+\Gamma ^i_{ni}
\Gamma ^n_{km}-\Gamma ^i_{mn}\Gamma ^n_{ki}, \label{2.4}
\end{equation}
where the notation $)_{,i}$ means take the partial derivative with respect
to $x^i$.  In turn, the Christoffel symbols are defined in terms of the
metric tensor as,
\begin{equation}
\Gamma ^m_{ij}={\textstyle {1\over 2}}g^{mk}(g_{ki,j}+g_{kj.i}-g_{ij,k}),
\;{\rm where}\; g^{mk}g_{kj}=\delta ^m_j, \label{2.5}
\end{equation}
defines $g^{km}$, and $\delta ^m_j$ is the Kronecker delta function.
Finally,
\begin{equation}
R=g^{ij}R_{ij}, \label{2.6}
\end{equation}
is the contraction of the Ricci tensor.

The $T^{44}$ element is the mass-energy
density. The $T^{4\beta}$ is the mass-flux density through an area
perpendicular to the direction $\beta $ per unit time.  (Greek indices
run from 1 to 3 while Roman indices run from 1 to 4.)  The $T^{\alpha \alpha }$
element is the pressure in the $\alpha $ direction, and the $T^{\alpha \beta}$
element is the flux density of the $\alpha $ component of momentum in the
$\beta $ direction. It is manifest that eq.~\ref{2.3} allows the direct
computation of the stress energy tensor from the metric tensor.  Tolman\cite{T}
gives the results for the form of the line element
\begin{equation}
ds^2=-e^\mu (dr^2+r^2d\theta ^2+r^2\sin ^2\theta d\phi ^2)+e^\nu dt^2,
\label{2.7}
\end{equation}
which is the spherical coordinate version or eq.~\ref{2.1}.  The
non-vanishing elements of the stress-energy tensor are
\begin{eqnarray}
8\pi T^1_1&=&-e^{-\mu}\left ({{\mu '^2}\over 4}+{{\mu '\nu '}\over 2}
+{{\mu '+\nu '}\over r}\right )\nonumber \\
&&+e^{-\nu}\left ( \ddot \mu +{\textstyle
{3\over 4}}\dot \mu^2-{{\dot \mu \dot \nu}\over 2}\right ) -\Lambda \nonumber \\
8\pi T^2_2&=&8\pi T^3_3=-e^{-\mu}\left ({{\mu ''}\over 2}+{{\nu ''}\over 2}
+{{\nu '^2}\over 4}+{{\mu '+\nu '}\over {2r}}\right ) \nonumber \\
&&+e^{-\nu }\left (
\ddot \mu +{\textstyle {3\over 4}}\dot \mu^2-{{\dot \mu \dot \nu}\over 2}
\right ) -\Lambda \nonumber \\
8\pi T^4_4&=&-e^{-\mu}\left (\mu ''+{{\mu '^2}\over 4}+{{2\mu '}\over r}
\right )+{\textstyle {3\over 4}}e^{-\nu}\dot \mu ^2 -\Lambda \label{2.8} \\
8\pi T^1_4&=&e^{-\mu}\left (\dot \mu '-{{\dot \mu\nu'}\over 2}\right )
\nonumber \\
8\pi T^4_1&=&-e^{-\nu}\left (\dot \mu '-{{\dot \mu\nu'}\over 2}\right )
\nonumber
\end{eqnarray}
where $\Lambda $ is the cosmological constant, an over dot denotes the
time derivative, and a prime denotes the derivative with respect to $r$.

Another quantity which is useful to consider is the extrinsic
curvature.\cite{MTW}  This concept arises in the Arnowitt {\it et al.}\cite{ADM}
splitting of four dimensional spacetime into 3 dimensional space plus
one dimensional time. The relationship between the metrics is given by
\begin{eqnarray}
ds^2&=&^{(4)}g_{ij}dx^idx^j \label{2.9} \\
&=&^{(3)}g_{\alpha\beta }(dx^\alpha +N^{\alpha }dt)
(dx^\beta +N^{\beta }dt)+N^2dt^2, \nonumber
\end{eqnarray}
where the $N^\alpha $ are the three shift functions and $N$ is the lapse
(of proper time) function. The intrinsic curvature is the analogue of
$R$ [eq.~\ref{2.6}] in three dimensions. The extrinsic curvature measures
the fractional shrinkage and deformation as one advances in time from
one space-like hyperplane to the next.  The extrinsic curvature tensor
is given as
\begin{equation}
K_{\alpha\beta}={1\over {2N}}\left [N_{\alpha |\beta }+N_{\beta |\alpha }
-{{\partial g_{\alpha \beta}}\over {\partial t}}\right ], \label{2.10}
\end{equation}
where the notation $)_{|\alpha }$ means the covariant derivative with respect
to $x^\alpha $.

The reason for the interest in the extrinsic curvature in our case is that
the necessary and sufficient junction conditions\cite{MTW} to join two
metrics in 4 dimensional spacetime across a 3 dimensional hypersurface is
that both $g_{\alpha\beta}$ and $K_{\alpha\beta}$ be continuous across
the surface.  In our case we are concerned with line elements of the
class of eq.~\ref{2.7}.  The split is between $r$ instead of $t$ and the
other three variables.  In this case the shift functions are all zero,
and the lag function is $N=\exp (0.5\mu )$.  The non-zero elements of $K$ are
\begin{eqnarray}
K_{22}&=&{\textstyle {1\over 2}}(\mu 'r^2+2r)\exp (0.5\mu ), \nonumber \\
K_{33}&=&{\textstyle {1\over 2}}(\mu 'r^2 +2r)\sin ^2\theta \exp (0.5\mu  ),
\label{2.11} \\
K_{44}&=&-{\textstyle {1\over 2}}\nu '\exp (\nu -0.5\mu) .\nonumber
\end{eqnarray}

Returning to the three-space, constant-time formalism, we give the
equations of motion of a free test particle as seen by local co-moving
observers along the test particle's path.  That is to say, a set of
observers whose coordinates do not change with time.  In other words,
we need the equations for the geodesic curves in spacetime. For the class of
line elements we are considering, it is simplest to start with a Lagrangian
formulation. By eq.~\ref{2.1} we may write this formulation as,
\begin{equation}
s=\int L\, dt, \Rightarrow L={{ds}\over {dt}}=
\left (g_{ij}\dot x^i\dot x^j\right )^{1/2}
\label {2.12}
\end{equation}
The standard Euler-Lagrange equations for an extreme in path length (Here
we seek a minimum distance between two fixed endpoints.) are
\begin{equation}
{d\over {dt}}\left ({{\partial L}\over {\partial \dot x^\alpha }}\right )
={{\partial L}\over {\partial x^\alpha }}. \label{2.13}
\end{equation}
Thus, using the diagonal nature of the metric tensor, we obtain for the
standard geodesic equations,
\begin{equation}
\left ({{ds}\over {dt}}\right ){d\over {dt}}\left [g_{\alpha i }
\left ({{ds}\over {dt}} \right )^{-1}\dot x^i \right ]
={1\over 2}g_{ij,\alpha}{{dx^i }\over
{dt}}{{dx^j}\over{dt}},  \label{2.14}
\end{equation}
where, of course $dx^4/dt =1$. From the line element we get
\begin{equation}
\left ({{ds}\over {dt}}\right )=\left [ e^\nu +g_{\alpha \beta}\dot x^\alpha
\dot x^\beta \right ]^{1/2} \label{2.15}
\end{equation}
These results display, for the class of line elements we are considering, the
three, second-order, non-linear, coupled equations
for the three coordinates $x^\alpha $ of a test particle as a function of time.
We have assumed isosynchronous coordinates  everywhere in the
three-dimensional, spacelike hyper-surface.  The square roots can removed
from eq.~\ref{2.14} by rewriting it as
\begin{eqnarray}
{1\over 2}\left ({{ds}\over {dt}}\right )^2{d\over {dt}}
\left [\left ({{ds}\over {dt}} \right )^{-2}\right ]
\left [g_{\alpha i }\dot x^i \right ]\nonumber \\  +
{d\over {dt}} \left [g_{\alpha i }\dot x^i \right ]
={1\over 2}g_{ij,\alpha}{{dx^i }\over
{dt}}{{dx^j}\over{dt}},  \label{2.16}
\end{eqnarray}

\section{``Swiss Cheese Model''}

As mentioned in the first section, there is a popular cosmological model
which in the large has the Friedmann-Lema\^\i tre line element,
\begin{eqnarray}
ds^2 &=&-{{a(t)^2}\over{[1+(r/2R)^2]^2}}[(dx^1)^2+(dx^2)^2+(dx^3)^2]\nonumber \\
&&+c^2(dx^4)^2 \label{3.1}
\end{eqnarray}
where in the notation of Sec. II,
\begin{eqnarray}
e^\mu&=&{{a(t)^2}\over{[1+(a(t)r/2a(t)R)^2]^2}} \nonumber \\
e^\nu &=&c^2. \label{3.2}
\end{eqnarray}
where $c$ is the velocity of light and $a(t)R$ is the radius of curvature of
the model universe.

The non-zero elements of the stress-energy tensor associated with this
line element are, by eq.~\ref{2.8}, 
\begin{eqnarray}
8\pi T^1_1&=&8\pi T^2_2=8\pi T^3_3={1\over {[a(t)R]^2}}+2{{\ddot a}\over{ac^2}}
+\left ({{\dot a}\over ac}\right )^2-\Lambda \nonumber \\
&&=-8\pi p_0, \label{3.3} \\
8\pi T^4_4&=&{3\over {[a(t)R]^2}}+3\left ({{\dot a}\over ac}\right )^2-\Lambda
=8\pi \rho _{00} \nonumber
\end{eqnarray}

The extrinsic curvature tensor, as given by eq.~\ref{2.11} for this metric has
the non-zero elements,
\begin{equation}
K_{22}={{a(t)r\left [1-(r/2R)^2\right ]}\over {[1+(r/2R)^2]^2}},\quad
K_{33}=\sin ^2\theta K_{22}. \label{3.4}
\end{equation}

In this cosmological model on scales small compared to that of the universe,
as in, for example, the solar system, a quite different metric is used.  It
is Schwarzschild's exterior solution.  The line element for it is
\begin{eqnarray}
ds^2&=&-\left (1+{m\over {2r}}\right )^4[(dx^1)^2+(dx^2)^2+(dx^3)^2]\nonumber \\
&&+ c^2\left ({{1-m/2r}\over {1+m/2r}}\right )^2(dx^4)^2, \label{3.5}
\end{eqnarray}
where in the notation of Sec. II, 
\begin{equation}
e^\mu =\left (1+{m\over {2r}}\right )^4, \qquad
 e^\nu =c^2\left ({{1-m/2r}\over {1+m/2r}}\right )^2. \label{3.6}
\end{equation}
Here $m$ is an abbreviation for $GM/c^2$ where $M$ is the mass concentration.
This metric is only valid outside the Schwarzschild radius $r>> r_0= 2GM/c^2$.
Inside this radius, Schwarzschild's interior solution is required but we
shall not be concerned with this aspect here.

 There are no non-zero elements of the stress-energy tensor associated with this
line element, unless the cosmological constant is different from zero.
Thus it corresponds to zero pressure and zero density, except
for a mass concentration in the center.  If $\Lambda \neq 0$ then,
\begin{equation}
T^1_1=T^2_2=T^3_3=T^4_4=-\Lambda /8\pi \label {3.7}
\end{equation}
This result would correspond to a uniform density and pressure through
out space, rather than the empty space with $\Lambda =0$.

A cross comparison of eq.~\ref {3.7} with eq.~\ref{3.3} shows that unless,
\begin{equation}
{1\over {[a(t)R]^2}} +\left ({{\dot a}\over ac}\right )^2= {{\ddot a}\over
{ac^2}} =0, \label{3.7.1}
\end{equation}
the stress-energy tensor is discontinuous at the boundary.  The only
solution of these equations, as $R$ is a constant, is $\dot a$ is a constant,
{\it i.e.}, a linear expansion factor, or a static flat universe.

Before we can us eq.~\ref{2.11} to compute extrinsic curvature for the
Schwarzschild case, we must first reparameterize the metric so that the
``Swiss cheese model'' metric boundary is given by one of the coordinates equals
a constant.\cite{DO} It will be convenient to use spherical coordinates.
If we follow the textbook \cite{Peb} approach, then we should start with
a zero pressure cosmological model.  By Birkoff's theorem\cite{Br} in this case
we may hollow out a sphere and replace it by a mass concentration at the center.
Since the density of the ``cosmic fluid'' is inversely proportional to $a^3$,
the radius of the sphere is fixed in the Friedmann-Lema\^\i tre coordinates.
The surface of the sphere is a three dimensional hypersurface parameterized
by the time and the two angle variables of spherical coordinates. To proceed,
we note that an observer sitting on the boundary is on a geodesic for
$r_{FL}=$ a constant for all time.  The relation between the Friedmann
Lema\^\i tre time $\bar t$ and the Schwarzschild variables as seen by
this observer is
\begin{eqnarray}
(ds)^2&=&c^2(d\bar t_e)^2\label {3.7a} \\
&=&\left [c^2\left ({{\displaystyle {1-{m\over {2r_e}}}}\over
{\displaystyle {1+{m\over {2r_e}}}}}\right )^2-\left (1+{m\over {2r_e}}
\right )^4 \left ({{dr_e}\over {dt_e}}\right )^2\right ](dt_e)^2, \nonumber
\end{eqnarray}
Since the observer is on a geodesic, we may deduce, by means of the time
component of eq.~\ref{2.14}, that he sees
\begin{equation}
c\left ({{\displaystyle 1 -{m\over {2r}}}\over{\displaystyle 1+{m\over {2r}}}}
\right )^2 {{dt}\over {ds}}=K^{1/2}, \label {3.7b}
\end{equation}
where $K$ is a constant of integration. Note is taken that, as the integration
is over $t$, $K$ may depend on $r$, of course.  By combining eq.~\ref{3.7a} and
eq.~\ref{3.7b}, we obtain,
\begin{eqnarray}
{{dr_e}\over {dt_e}}&=&{{\displaystyle c\left (1-{m\over {2r_e}}\right )}\over
{\displaystyle \left (1+{m\over {2r_e}}\right )^3}}\left [
1-{1\over K}\left ({{\displaystyle 1-{m\over {2r_e}}}\over
{\displaystyle 1+{m\over {2r_e}}}}\right )^2\right ]^{1/2}, \nonumber \\
{{dt_e}\over{d\bar t_e}}&=&K^{1/2}\left ({{\displaystyle 1+{m\over {2r_e}}}\over
{\displaystyle 1-{m\over {2r_e}}}}\right )^2,\label {3.7c} \\
{{dr_e}\over {d\bar t_e}}
&=&{{cK^{1/2}}\over{\displaystyle \left (1-{{m^2}\over {4r_e^2}}\right )}}
\left [ 1-{1\over K}\left ({{\displaystyle 1-{m\over {2r_e}}}\over
{\displaystyle 1+{m\over {2r_e}}}}\right )^2\right ]^{1/2}, \nonumber
\end{eqnarray}
which gives the behavior of $r_e,\; t_e$ as a function of the
Friedmann-Lem\^\i tre time variable, as seen by the comoving observer
sitting on the metric boundary in the Swiss-cheese model.

By equating the coefficients of the angular variables, we obtain,
\begin{eqnarray}
\left (1+{m\over {2r_e}}\right )^2r_e&=&{{a(\bar t_e)\bar r_e}\over
{\displaystyle 1+\left ({{\bar r_e}\over {2R}}\right )^2}}, \label{3.7d} \\
{{\dot a(\bar t_e)}\over {a(\bar t_e)}}&=&{1\over r_e}{{dr_e}\over {dt_e}}
\left ({{\displaystyle 1-{m\over {2r_e}}}\over {\displaystyle 1+{m\over
{2r_e}}}} \right ){{dt_e}\over {d \bar t_e}} \nonumber
\end{eqnarray}
Note is taken that the boundary coordinate $\bar r_e$ in the
Friedmann-Lema\^\i tre metric depends on the central mass concentration,
and on the radius of curvature of the universe, {\it i.e.} on $m$ and $R$.

Next we combine eqs.~\ref{3.7c} and \ref{3.7d} and comparing the result
with the ``cosmological equation.'' This equation is the equation for the
$T^4_4$ component of the stress energy tensor derived from the
Friedmann-Lema\^ \i tre line element.  The result of the application of
this textbook method is that the constant of integration is
\begin{equation}
K=\left [{{\displaystyle 1-\left ({{\bar r_e}\over {2R}}\right )^2}\over
{\displaystyle 1+\left ({{\bar r_e}\over {2R}}\right )^2}}\right ]^2,
\label {3.7d1} \end{equation}
which is a function of $m$ and $R$.

We are now in a position to introduce a reparameterization of the Schwarzschild
metric.  We choose the new parameters $\hat t =\bar t$ and $\hat r$.  The
second variable is chosen so as to keep the metric continuous at the
interface, and to keep the metric diagonal, if possible.  The reparameterized
Schwarzschild metric is given by eq.~\ref{5.0}.  Eq.~\ref{3.7a} insures
that $\hat g_{44}=c^2$.  The equation of continuity, and the vanishing of
the elements $\hat g_{14}=\hat g_{41}$ yield the conditions,
\begin{eqnarray}
\left ({{dr_e}\over {d\hat r_e}}\right )^2g_{11}+\left ({{dt_e}\over
{d\hat r_e}} \right )^2g_{44}&=&\bar g_{11},\nonumber \\
\left (g_{11}{{dr}\over {d\hat t}}\right )
{{dr}\over{d\hat r}}+\left ({{dt}\over {d\hat t}}g_{44}\right ){{dt}\over
{d\hat r}}&=&0 \label{3.7e}
\end{eqnarray}
The solution of these equations is
\begin{eqnarray}
{{dr}\over {d\hat r}}&=&{{K^{1/2}a(\hat t)}
\over{\displaystyle \left (1-{{m^2}\over {4r^2_e}}\right )\left [
1+\left ({{\bar r_e}\over {2R}}\right )^2\right ] }},\nonumber \\
{{dt}\over {d\hat r}}&=&
{{\displaystyle K^{1/2} a(\hat t)\left (1+{m\over {2r}}\right )^3}
\over{\displaystyle c\left [1+\left ({{\bar r_e}\over {2R}}\right )^2\right ]
\left ( 1-{{m^2}\over {4r^2_e}}\right)\left (1-{{m}\over {2r}}\right )}}
\nonumber \\ &&\times \left [1-{1\over K}
\left ({{\displaystyle 1-{m\over {2r}}}\over {\displaystyle 1+{m\over {2r}}}}
\right )^2 \right ]^{1/2}, \label {3.7f} \\
\hat g_{11}&=&-{{\displaystyle \left (1-{{m^2}\over {4r^2}}\right )^2
a(\hat t)^2 }\over{\displaystyle \left (1-{{m^2}\over {4r^2_e}}\right )^2
\left [1+\left ({{r_e}\over {2R}}\right )^2\right ]^2}} \nonumber
\end{eqnarray}

Our construction insures the continuity of $\hat g_{11}=\bar g_{11}$ at the
metric interface.  The continuity of the (22) and the (33) elements are insured
by eq.~\ref{3.7d}. All the off-diagonal elements vanish.

We may now apply eq.~\ref{2.11} to obtain the extrinsic curvature.  The
result is just eq.~\ref{3.4} with $(r,\; t,\; \theta)$ replaced by
$(\hat r,\; \hat t,\; \hat\theta )$. This result shows the continuity of
the extrinsic curvature.

There is however, one important item to note.  Consider the derivatives
perpendicular to the boundary hypersurface, evaluated at the interface.
\begin{eqnarray}
{{d\log (-\bar g_{11})}\over {d\bar r}}&=& -{{\bar r}\over {R^2+0.25\bar r^2}},
\nonumber \\
{{d\log ( -\hat g_{11})}\over {d\hat r}} &=&
-{{\displaystyle a(\hat t){m\over {r_e^3}}\left [1-\left ({{r_e}\over {2R}}
\right )^2\right ]}\over {\displaystyle\left (1-{{m^2}\over{4r_e^2}}\right )^3
\left [ 1+\left ({{r_e}\over {2R}}\right )^2\right ]^2}}. \label {3.8}
\end{eqnarray}
To leading order in a flat universe, eq.~\ref{3.8} becomes,
\begin{equation}
{{d\log (-\bar g_{11})}\over {d\bar r}}=0,
\qquad {{d\log ( -\hat g_{11})} \over {d\hat r}}
=-a(\hat t){m\over {r_e^3}}. \label {3.9}
\end{equation}
That such a discontinuity may occur is well known\cite {Stef}.
I will explore some of the consequences of this discontinuity in the
next section.

As was remarked above,
it is well known that it is a necessary condition that the pressure equal zero
in order for this matching to occur\cite{Peb}.  As the pressure $p=-T_\alpha
^\alpha $ for each $\alpha $ (not summed here), the zero-pressure condition
means, by eq.~\ref{3.7} which gives the pressure for the Schwarzschild case
in terms of the $T_\alpha ^\alpha $ elements that $\Lambda =0$. Turning to
the the Friedmann-Lema\^\i tre case, we see from eq.~\ref{3.3} that it must
be that
\begin{equation}
{1\over {[a(t)R]^2}}+2{{\ddot a}\over{ac^2}}
+\left ({{\dot a}\over ac}\right )^2=0. \label {3.10}
\end{equation}
If use substitute the standard form $a(t)=(t/t_0)^\psi$ in eq.~\ref{3.10},
then the only solutions are $R=\infty $ together with $\psi =0$ or $2/3$.
The first solution is the trivial case of a static universe and is of no
interest in the present discussion. In the second case, the stress-energy
tensor element $T_4^4>0$, which means there is a mass discontinuity. This
is expected by the structure of the approximation of scooping out a hollow
sphere, and then having a uniform density outside.

\section{Dynamics in the ``Swiss Cheese Model''}

Following up on the discontinuity in the derivate of the metric, perpendicular
to the metric boundary in the ``Swiss Cheese Model,'' we investigate
the some of the dynamic properties of this model. It suffices for my
purposes to consider the simpler case of a flat ($R=\infty $) universe.
The equations of motion are as follows.  For the Schwarzschild metric,
we get from eq.~\ref{2.14}, to leading order, the well known Newtonian equations
\begin{equation}
{{d^2\vec \rho}\over {dt^2}}=-{{GM}\over{\rho ^3}}\vec \rho \label {6.01}
\end{equation}
The solutions to this equation are the familiar Newtonian conic sections.

For the Friedmann-Lema\^\i tre metric from eq.~\ref{2.14} we get a first
intergal as
\begin{equation}
\dot {\vec x} ={{\vec A}\over{a^2(t)}} \label {6.02}
\end{equation}
For my purposes, it is more convenient to use a variable more closely
equal to the proper distances.  Thus for $\vec \rho =a(t)\vec x$,
eq.~\ref{6.02} becomes,
\begin{equation}
{{d^2\vec \rho }\over {dt^2}}={{\ddot a(t)}\over {a(t)}}\vec \rho
\label {6.03}
\end{equation}
In both metrics $\rho $ is within plotting accuracy for the proper distance
in the examples I will consider.

The general solution of eq.~\ref{6.03} is
\begin{equation}
\vec \rho =\vec At^\psi +\vec Bt^{1-\psi}= \vec At^{2/3}+\vec Bt^{1/3},
\label{6.3}
\end{equation}
when the currently, theoretically favored value of $\psi ={2\over 3}$ is
chosen. Let us take the example in rectangular coordinates where
initially $x=\lambda ,\;\dot x =0,\; y=0,\; \dot y =\lambda $ at time $t=t_0$.
Then eq.~\ref{6.3} becomes,
\begin{eqnarray}
x&=&\lambda \left [-\left ({t\over {t_0}}\right )^{2/3}+2\left ({t\over {t_0}}
\right )^{1/3}\right ],\nonumber  \\
 y&=&3t_0\lambda \left [\left ({t\over {t_0}}\right )^{2/3}-
\left ({t\over {t_0}}\right )^{1/3}\right ] . \label{6.3a}
\end{eqnarray}
It is worth noting that for $t-t_0\ll t_0$ that the motion is almost
exactly that of a straight line, as is to be expected in flat empty
space.  Specifically, direct calculation yields
\begin{equation}
\ddot x(t_0)= -{2\lambda \over {9t_0^2}}, \qquad \ddot y(t_0) = 0.
\label {6.3b} \end{equation}
This apparent acceleration is of order $H_0^2$.

The solutions in eq.~\ref{6.3} are quadratic in the parameter $t^{1/3}$.
Thus we can obtain the equation for the trajectory as a quadratic plus
linear expression in $\rho _x$ and $\rho _y$. The form would be
\begin{equation}
\rho _y=c(a\rho _x+b\rho _y)^2 +d(a\rho _x+b\rho _y) \label{6.4}
\end{equation}
This form is readily recognized as a parabola.  Thus it is the case
that a freely moving particle in an Friedmann-Lema\^\i tre expanding
space always appears to be moving in a parabola.  This effect is caused
by the small $\ddot a$ term which appears in the equation \ref{6.03} as
a forcing term.

As an illustration of the behavior of the ``Swiss cheese model''
I have computed the following trajectories.  I use as
a unit of time the Hubble time, that is $1/H_0$, which is of the order of
$10^{18}$ seconds.  As a unit of distance I use $\root 3 \of{M_\odot G/H_0^2}$
which is about 20 million Astronomical units. $M_\odot $ is the mass of the
sun.  I set $t_0=1$ to switch to our current units.  The metric interface
is at $r=t^{2/3}$ in these units for flat spacetime, as mentioned above.
In order to follow a Friedmann-Lema\^\i tre trajectory the test particle's
distance from the origin must be larger at every time than that for the
interface.  Thus,
\begin{equation}
\lambda ^2\left [10t^{4/3}-22t +13t^{2/3}\right ] > t^{4/3}, \label{6.11a}
\end{equation}
The trajectory will intersect the interface if eq.~\ref{6.11a} is an equality.
By means of the quadratic formula, an intersection will occur if
\begin{equation}
t^{1/3}={{11\pm\sqrt{-9 +13/\lambda ^2}}\over {10 -1/\lambda ^2}}.
\label{6.11b}
\end{equation}
It will be observed that for $\lambda < \sqrt{13}/3 $ there are two
real roots.  If $\lambda =1$, then $t^{1/3}=1,\; 13/9$.  On the other
hand, if $\lambda > \sqrt{13}/3$, the roots are imaginary, so there are
no intersections. If $\lambda =\sqrt{13}/3$ there is a double root at
$t^{1/3}=13/11$.  In this case the parabolic trajectory just grazes the
metric interface.

\begin{figure}
\vskip 0.5\baselineskip
\centerline {\psfig{figure=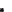,width=\hsize}}
\vskip 0.5\baselineskip
\figure{FIG.\ 1\ \ ``Swiss cheese model'' trajectories which begin at
$x_0=\sqrt{13}/3 +\epsilon $ (the Friedman-Lema\^\i tre case), and
begin at $x_0=\sqrt{13}/3 -\epsilon $ (the Schwarzschild case). Here,
$\epsilon >0$ may be chosen as small as one pleases.  The initial part
of both trajectories is a parabola generated by the Friedmann-Lema\^\i tre
metric.  At time $t = (13/11)^3$, marked in the figure, the two trajectories
separate.\label{fig.1}}
\end{figure}

In Fig.~1, I illustrate the two different trajectories when $r_0=
\sqrt{13}/3$ and $\dot \phi _0=1.0$. In the case where $r_0$ is just
any arbitrary amount smaller, the expanding interface overtakes the
test particle following its Friedmann-Lema\^\i tre parabolic trajectory
and it must then follow the static Schwarzschild equations of motion.
The Schwarzschild metric takes over at $t=(13/11)^3$ as explained above,
and after that the trajectory is an ellipse with
semimajor axis 7.6287 and the semiminor axis 3.986.  These imply an
eccentricity of 0.69068 and a semilatus rectum parameter of $p= 2.086$.
On the other hand, if $r_0$ is any arbitrary amount larger, it escapes
the moving interface and continues to follow the parabolic trajectory.
It is evident from Fig.~1 that future trajectories are, in some cases,
discontinuous functions of the initial conditions for the ``Swiss
cheese model.''

Put another way, the Schwarzschild metric permits closed orbits and
the Friedmann-Lema\^\i tre metric does not.  In terms of the latter
coordinates, one can choose initial conditions so that the parabolic
trajectory just grazes the metric interface (fixed radial coordinate
in this metric) and the speed is low enough that for infinitesimally
different initial conditions the trajectory crosses the interface and
is caught in a bound state, or alternatively misses the interface and
proceeds on its parabolic trajectory.  All these effects take place in
supposedly empty space of the order of 20 million AU from a mass concentration
of size $M_\odot $, and are quite counter to one's physical intuition
that such discontinuities should not occur there.

\section{An Acceptable Metric}

I propose the following line element to represent a mass condensation in
an expanding and curved universe.  It is of the form \ref{2.7} where
\begin{eqnarray}
e^\mu &=&{{a(t)^2}\over {[1+(a(t)r/2a(t)R)^2]^2}}\left (1+{m\over {2a(t)r}}
\right )^4,\label{4.1} \\
e^\nu &=&c^2\left ({{1-m/2a(t)r}\over {1+m/2a(t)r}}\right )^2 ,\quad
m\equiv {{GM}\over {c^2}}\left [1+\left ({r\over {2R}}\right )^2\right ]
^{1/2} \nonumber
\end{eqnarray}
where $G$ is Newton's constant of gravitation.  It is to be noted that
this metric is an adaptation of the Schwarzschild metric in curved space.
It is not claimed that this metric is unique. Certainly any coordinate
transformation of this metric is equivalent. It does however have the property
that in the limit where $a(t)$ is a constant, it reduces to the
Schwarzschild metric in curved space.  Also, when the central mass
vanishes it reduces to the Friedmann-Lema\^\i tre metric.  It is
an infinitely differentiable solutions to the Einstein
field equations which corresponded to values of the stress-energy tensor
which are only of second order in $R^{-1}$ and $H_0$, except at the central
mass. This latter property is also true of the Friedmann-Lema\^\i tre metric
in curved space, as may be seen in eq.~\ref{3.3}.  Thus we have, by this example
whose properties are globally similar to the Friedmann-Lema\^\i tre metric,
demonstrated that the Swiss cheese model is not require to match the expansion
of the universe observed at large scales and the absence of any such expansion
observable in the solar system.  As shown in the previous section since
the Swiss cheese model has a rather severe and very unphysical deficiency,
we think it is time to begin the search for an acceptable metric.

The non-zero elements of the extrinsic curvature is given by (2.11) as
\begin{eqnarray}
K_{22}&=&a(t)r {{[1-(r/2R)^2]}\over{[1+(r/2R)^2]^2}}[1-(m/2a(t)r)^2]
 \nonumber \\
K_{33}&=&\sin^2\theta K_{22} \label{4.2} \\
K_{44}&=&-{{mc^2(1-m/2a(t)r)}\over {a(t)^2r^2(1+m/2a(t)r)^5}} \nonumber
\end{eqnarray}
In the limit that $m\to 0$ these curvatures reduce to those of eq.~\ref{3.4} and
in the limit $R\to \infty $ and $\dot a =0$ ($a=1$) they reduce to
\begin{eqnarray}
K_{22}&=&r\left [1-\left ({m\over {2r}}\right )^2\right ],\quad
K_{33}=\sin^2\theta K_{22}, \nonumber \\
K_{44}&=&-{{mc^2(1-m/2r)}\over {r^2(1+m/2r)^5}} \label{4.2a}
\end{eqnarray}
These curvatures agree with those of the un\-re\-pa\-ram\-e\-ter\-ized
Schwarzschild metric. The curvatures in eq.~\ref{4.2} differ only in that
$r$ is replaced by $a(t)r$, $R$ by $a(t)R$, and there are corrections for the
overall curvature of space.  Thus, it reflects, as does the metric, the
very same behavior, in terms of $a(t)r$ instead of $r$ as was found by
Schwarzschild for his metric.

Both the metric and the extrinsic curvature are continuous outside the
Schwarzschild radius, which is necessary for a metric to be acceptable.

We now turn to the stress-energy tensor.  The current metric shares with
both the Friedmann-Lema\^\i tre metric, eq.~\ref{3.1}, and the Schwarzschild
metric, eq.~\ref{3.5}, the property that $T_1^4=T_4^1=0$, so there is no
mass-density flux. This property follows by direct computation from
eq.~\ref{2.8} and eq.~\ref{4.1}.  For the other non-zero components, we
find
\begin{eqnarray}
8\pi T^1_1&=& {{1+(m/2a(t)r))^2}\over {a(t)^2R^2(1+m/2ar)^5
(1-m/2a(t)r)}}\nonumber \\
&&+{{2\ddot aa[1+m/2a(t)r)]+3\dot a^2[1 -m/2a(t)r]
}\over{a(t)^2c^2(1-m/2a(t)r)}} -\Lambda \nonumber \\
8\pi T^2_2&=&8\pi T^3_3= 8\pi T^1_1  \label {4.3} \\
8\pi T_4^4&=& {3\over {a(t)^2R^2(1+ m/2a(t)r)^5}}
+{{3\dot a^2}\over {c^2a^2}}-\Lambda , \nonumber
\end{eqnarray}
In the limit as $m\to 0$ or $r\to \infty $ these tensor elements reduce to
those of eq.~\ref{3.3},
and also in the limit as $R\to \infty $ and $\dot a\to 0$ they vanish as with
the Schwarzschild metric, unless $\Lambda \neq 0$, in which case the
result (3.7) is obtained.  The dominate terms in the diagonal
elements (pressure and density) are a sum of terms of second order in
$1/R$ and terms containing two time derivatives of the universal expansion
factor $a(t)$. The metric given by eqs.~\ref{4.1} and \ref{2.7} is the
solution of Einstein's field equations eq.~\ref{2.3} when the stress-energy
tensor, eq.~\ref{4.3}, is specified.

\section{Dynamics}

In this section we investigate the equations of motion a test particle in
our proposed metric as described by eq.~\ref{2.7} and eq.~\ref{4.1}.
Our treatment diverges from that of McVittie\cite{McV} at this point as
he directly substitutes $\rho =a(t) r$ into the line element rather than
using the transformation equations 
\begin{equation}
\bar g_{kl}={{\partial x^i}\over {\partial \bar x^k}}
{{\partial x^j}\over {\partial \bar x^l}}g_{ij}, \label{5.0}
\end{equation}
The correct change of the line element for this change of variables is
\begin{eqnarray}
ds^2&=&-e^U\left [(d\rho )^2+\rho ^2(d\theta )^2+\rho ^2\sin ^2 \theta
(d\phi )^2\right ]\nonumber \\ && +\left [e^V -e^U\left ( {{\rho \dot a}
\over {a}}\right )^2 \right ](d\tau )^2 +2e^U\left ( {{\rho \dot a}\over {a}}
\right )d\rho d\tau ,
\label{5.01}
\end{eqnarray}
where
\begin{eqnarray}
e^U&=&{{(1+m/2\rho )^4}\over {[1+(\rho /2a(\tau )R)^2]^2}} \nonumber \\
e^V &=&c^2\left ( {{1-m/2\rho }\over {1+m/2\rho }}\right )^2 \label{5.02}
\end{eqnarray}
Clearly these are non-synchronous coordinates as the coefficient of
$d\rho d\tau $ is non-zero.

We will make this change of variables later after the equations of motion
have been derived.  The main difference is that McVittie omits the $\dot
\mu $ terms from his subsequent equations. The equations of motion then
become, using eq.~\ref{2.14}, for $\theta $ and $\phi $,
\begin{eqnarray}
\left({{ds}\over {dt}}\right ){d\over {dt}}\left [\left ({{ds}\over {dt}}
\right )^{-1}e^\mu r^2\dot \theta \right ]
&=&e^\mu r^2\sin \theta \cos \theta \dot \phi ^2 \nonumber \\
\left({{ds}\over {dt}}\right ){d\over {dt}}\left [\left ({{ds}\over {dt}}
\right ) ^{-1} e^\mu r^2\sin ^2\theta \dot \phi\right ]&=& 0 \label{5.1}
\end{eqnarray}
One can see by inspection, the motion in the plane $\theta =\pi /2$ is
a solution.  The intergal of the second equation gives the result,
\begin{equation}
r^2\dot \phi = Ae^{-\mu }\left ({{ds}\over {dt}}\right ), \label {5.2}
\end{equation}
where $A$ is a constant of integration.  This equation is the conservation
of angular momentum in this coordinate system.  To obtain the equation
of motion for $r$, it is convenient to use eq.~\ref{2.16}. We obtain,
\begin{eqnarray}
-{1\over 2}\left ({{ds}\over {dt}}\right )^2{d\over {dt}}
\left [\left ({{ds}\over {dt}} \right )^{-2}\right ]e^\mu\dot r
-{d\over {dt}}\left [e^\mu \dot r\right ]= \nonumber \\
-{1\over 2}\Big [\mu 'e^\mu
\dot r^2+(\mu'r^2+2r)e^\mu \dot \theta ^2 \nonumber \\
+(\mu'r^2+2r)e^\mu\sin ^2\theta
\dot \phi ^2-\nu 'e^\nu \Big ],\label{5.3}
\end{eqnarray}
where, from the line element,
\begin{equation}
\left ({{ds}\over {dt}}\right )^2=-e^\mu\left [\dot r^2+r^2\dot\theta ^2
+r^2\sin ^2 \theta \dot\phi ^2\right ] +e^\nu . \label{5.4}
\end{equation}

At this point, we simplify to motion in the $\theta = \pi /2$ plane. Thus
$\dot \theta =\ddot \theta =0$, and eq.~\ref{5.3} becomes,
\begin{eqnarray}
{1\over 2}\left ({{ds}\over {dt}}\right )^2{d\over {dt}}
\left [\left ({{ds}\over {dt}} \right )^{-2}\right ]\dot r
+e^{-\mu }{d\over {dt}}\left [e^\mu \dot r\right ]= \nonumber \\
{1\over 2}\left [\mu ' \dot r^2+(\mu'r^2+2r)
\dot \phi ^2-\nu 'e^{\nu -\mu}\right ].\label{5.5}
\end{eqnarray}

These formulas yield the equations of motion of a test particle in the
reference frame which is at rest with respect to the coordinate system
at the location of the test particle.  We want to find the motion with
respect to an observer at rest at $r=0$. To this end we introduce the
change of variables,
\begin{equation}
r={\rho \over {a(t)}},\qquad \dot r={{\dot \rho }\over {a(t)}}
-{{\dot a(t)\rho }\over{a(t)^2}} \label {5.6}
\end{equation}
This change of variables yields coordinates which are equal to
the proper distances, as viewed from the origin, when the mass concentration
is absent. With the substitution \ref{5.6} we obtain for eq.~\ref{5.2}
and eq.~\ref{5.4}
\begin{eqnarray}
\rho^2\dot \phi &=& Aa(t)^2e^{-\mu }\left ({{ds}\over {dt}}\right ),
\label {5.7}\\
\left ({{ds}\over {dt}}\right )^2&=&-{{e^\mu}\over {a(t)^2}}
\left [\left (\dot \rho -{{\dot a(t)\rho }\over{a(t)}}\right )^2
+\rho ^2 \dot\phi ^2\right ] +e^\nu . \label {5.8}
\end{eqnarray}
Finally, eq.~\ref{5.5} becomes,
\begin{eqnarray}
&&\ddot \rho -2{{\dot a}\over a}\dot \rho +\rho \left [2\left ({{\dot a}
\over a}\right )^2-{{\ddot a }\over a}\right ]+\dot \mu\left (\dot \rho
-{{\dot a}\over a}\rho \right ) \label{5.9} \\
&=&{1\over 2}\mu 'a\left ({{\dot \rho }
\over a}-{{\dot a}\over {a^2}}\rho \right )^2+{1\over 2}\left (
\mu '{{\rho }\over a}
+2\right ) \left [{{A^2a^4e^{-2\mu}}\over {\rho ^3}}\right ]\left ({{ds}
\over {dt}}\right )^2 \nonumber \\
&&-{1\over 2}\nu 'a e^{\nu -\mu}
-{1\over 2}\left ({{ds}\over {dt}}\right )^2{d\over {dt}}
\left [\left ({{ds}\over {dt}} \right )^{-2}\right ]\left (\dot \rho
-{{\dot a}\over a}\rho \right ) \nonumber
\end{eqnarray}
where we have used eq.~\ref{5.7} to eliminate the $\dot \phi $ dependence.
There is also a $\dot \phi $  in $ds/dt $ but it too can be eliminated by
the substitution of $\dot \phi $ from eq.~\ref{5.7} in eq.~\ref{5.8}.
To assess the importance of the various terms it is helpful to note the
following dimensionless quantities, in ``planetary units,''
\begin{eqnarray}
T_{\oplus}H_0&\approx &5\times 10 ^{-11},\quad \left ({{v_{\oplus}}\over c}
\right )^2\approx 1\times 10 ^{-8},\nonumber \\ {{GM_\odot}\over {c^2R_\oplus }}
&\approx &1\times 10 ^{-8},\quad \left ({{R_\oplus }\over {R_{\rm Hubble }}}
\right )^2 \approx 0.6\times 10 ^{-30}, \label{5.9z}
\end{eqnarray}
where $T_\oplus,\;v_\oplus,\; R_\oplus $ are the orbital period, velocity,
and radius of the earth, and $M_\odot$ is the mass of the sun.
Some limiting cases are of interest.  First,  we take the flat-space,
slow-speed limit, {\it i.e.}, $R=\infty $, and
\begin{equation}
\left ({{ds}\over {dt}}\right )^2\approx c^2\left ({{\displaystyle 1-{{GM}
\over {2c^2\rho }}}\over {\displaystyle 1+{{GM}\over {2c^2\rho }}}}\right )
^2 . \label{5.9a}
\end{equation}
Thus eq.~\ref{5.9} reduces to
\begin{eqnarray}
&&\ddot \rho -{{\ddot a}\over a}\rho -\left (\dot \rho - {{\dot a}\over a}\rho
\right ){{\dot am(3-m/\rho )}\over {a\rho \left [1-(m/2\rho )^2\right ]}}
\label {5.9b} \\
&=&-{{m(\dot \rho -\dot a\rho /a)^2}\over {\rho ^2(1+m/2\rho )}}
+{{A^2c^2(1-m/2\rho )^3}\over{\rho ^3(1+m/2\rho )^{11}}}-
{{mc^2(1-m/2\rho )}\over {\rho ^2(1+m/2\rho )^7}} \nonumber
\end{eqnarray}
We may further reduce these equations by discarding terms which are proportional
to $c^{-2}$ for when the velocities are much less than the speed of light.
By eq.~\ref{4.1} these are the terms proportional to m alone, but we retain,
of course, the terms in $mc^2$. Eq.~\ref{5.9b} reduces further to
\begin{equation}
\ddot \rho -{{\ddot a}\over a}\rho ={{A^2c^2}\over {\rho ^3}}-{{GM}\over
{\rho ^2}} ={{GM}\over {\rho ^2}}\left ( {{\rho _0}
\over \rho } -1 \right ).  \label {5.10}
\end{equation}
The coefficient of the discarded term on the left-hand side of eq.~\ref{5.9b}
is of the order of $10^{-19}$ as it is the product of two first order
corrections. The discarded (first) term on the right-hand side of
eq.~\ref{5.9b} is smaller by a factor of $\dot \rho ^2/c^2 $ than
the last term.  The other items discard are factors of $GM/c^2\rho $ smaller
than the dominant terms.

It is to be noticed that in the limit of eq.~\ref{5.10}, that it differs from
Newton's equation of gravitation only by a term proportional to $\ddot a$.
Note is taken the Noerdlinger and Petrosian\cite{NP} do take account of this
term.  The constant $\rho _0$ is just another form of the constant of
integration $A$. The magnitude of the $\ddot a$ term equals that for
the sun's gravity at about 0.5 kiloparsecs.  The effects on the scale
of the solar systems are too small to be measured.

\section{Examples}

The equations of motion in a flat, Friedmann-Lema\^\i tre expanding universe
for a slowly moving test particle under
no external forces are, by eq.~\ref{5.10} and the corresponding reduction
of eq.~\ref{5.1},
\begin{equation}
\ddot \rho -\rho \dot \phi ^2={{\ddot a}\over a}\rho
=-{{\psi (1-\psi )}\over {t^2}}\rho ,\qquad {d\over {dt}}\left (\rho ^2
\dot \phi \right )=0, \label{6.1}
\end{equation}
for the standard form of the universal expansion factor for the universe,
as describe at the end of section III. $t$ is the current age of the
universe. One easily recognizes these equations
to be just exactly Newton's equations in a plane in spherical coordinates.
If we change to rectangular coordinates, we get exactly eq.~\ref{6.03}.
The behavior of the solutions is discussed in detail in Section IV above.

Next we consider the case where we add gravitational effects to their
leading order. For purely radial motion, the $A$ of eq.~\ref{5.7} is zero.
Thus the equation of motion \ref{5.10} becomes,
\begin{equation}
\ddot {\rho} ={{\ddot a(t)}\over {a(t)}}\rho -{{GM}\over {\rho ^2}}
\label {6.5}
\end{equation}
This equation differs from the Schwarzschild metric equation, \ref{6.01}
by the addition of a term in $\ddot a$.
We will consider this behavior over time periods short compared to the age
of the universe. If we multiply by $d\rho $ and integrate, we get,
\begin{equation}
{1\over 2}\dot \rho ^2={{\ddot a}\over {2a}}\rho ^2 +{{GM}\over {\rho}} +E_0
\label {6.6} \end{equation}
where $E_0$ is the constant of integration. If $\ddot a=0$, then $E_0=0$
would correspond to a test particle which had zero velocity at $\rho =\infty $.
I choose to examine this special case. Then eq.~\ref{6.6} becomes
\begin{eqnarray}
\int _{\rho _0}^\rho {{\xi ^{1/2}d\xi }\over {2\sqrt {GM+{{\ddot a}\over {2a}}
\xi ^3}}}&=& t-t_0. \quad \Rightarrow \nonumber \\
\left .{2\over 3}\sqrt {-{a\over{
\ddot a}}} \sin^{-1}\left ( {{\xi ^{3/2}}\over {\sqrt {-2GMa/\ddot a}}}\right )
\right |_{\rho _0} ^{\rho } &=&t-t_0, \label{6.7}
\end{eqnarray}
by Pierce's tables\cite{Pie}.  Thus,
\begin{eqnarray}
\rho ^{3/2}&=&\rho _0^{3/2}+\sqrt{-{{2GMa}\over {\ddot a}}}\left [
\sin \left ({3\over 2}(t-t_0)\sqrt {-{{\ddot a}\over a}}\right ) \right ]
\label {6.8}  \\
&\approx &\rho _0^{3/2}+{3\over 2} \sqrt {2GM}\left [ (t-t_0) +{{3\ddot a}\over
{8a}}(t-t_0)^3 + \ldots \right ].  \nonumber
\end{eqnarray}
It is evident from this solution that the leading order corrections due
to the expansion of the universe are of the order $H_0^2(t-t_0)^2$ which
is extremely small on the planetary time scale.  The solution when $(t-t_0)
=O(t_0)$ would take account of the time dependence of $\ddot a/a$.  The more
general case, where $E_0\neq 0$, can also be integrated in terms of elliptic
functions of the first and third kinds\cite{GR}.

Next we investigate bound circular motion.  To do so, I set $\ddot \rho =0$
in eq.~\ref{5.10} which gives,
\begin{equation}
{{A^2c^2}\over {\rho ^3}}= {{GM}\over {\rho ^2}}+{{\ddot a}\over a}\rho .
\label{6.9} \end{equation}
If I use eq.~\ref{5.7} to reintroduce $\dot \phi $, and remember that the
period $T=2\pi/\dot \phi $, then I find,
\begin{equation}
{{4\pi ^2\rho ^3}\over {GMT^2}}=1+{{\ddot a\rho ^3}\over {GMa}}, \label{6.10}
\end{equation}
which is Kepler's law relating the square of the period to the cube of
the radius, with a correction caused by the expansion of space.  For
the case $a(t)\propto t^{2/3}$ the correction term becomes, $-{1\over 2}H_0^2
\rho ^3/GM$.  If we use the solar mass, then
\begin{equation}
{{4\pi ^2\rho ^3}\over {GMT^2}}=1-3.307\times 10 ^{-23}h^2_{50}\rho ^3 ,
\label{6.11} \end{equation}
where $\rho $ is in astronomical units and $h_{50}=1$ when $H_0=50$ km per
second per megaparsec.

\begin{figure}
\vskip 0.5\baselineskip
\centerline {\psfig{figure=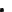,width=\hsize}}
\vskip 0.5\baselineskip
\figure{FIG.\ 2\ \ Trajectories in the static (Schwarzschild) metric and
the non-static (Friedman-Lema\^\i tre) portions of the ``Swiss cheese
model.''  The trajectory for the metric of this paper, with the same initial
conditions as for the Schwarzschild case, is also shown. \label{fig.2}}
\end{figure}

These results are illustrated in Fig.~2.
In flat space at time $t_0$, the interface
between the static Schwarzschild metric and the non-static
Friedmann-Lema\^\i tre
metric is at distance unity from the origin, in the units of section IV.
We display the Schwarzschild result when started with unit angular velocity
($\dot \phi =1$). It is a circle of radius unity. Next we display the
trajectory just outside the interface.  It is, as expected, a parabolic
curve.  I have started it with unit angular velocity (again $\dot \phi =1$)
at a distance of 1.25 from the mass concentration. It is to be noted that
after a unit (Hubble) time has passed, the two trajectories are
significantly separated.

The reason for starting
it further out is, as explained in section IV, that if I were to have started
it at a distance between
1.0 and $\sqrt{13}/3\approx 1.2018504 $, the trajectory would have been
over taken by the expanding spherical interface between the two metrics
of the ``Swiss cheese model.''

In addition I show the trajectory using the presently considered  metric.
I have again started with unit distance and unit angular velocity.  The
result here is that it converges fairly quickly to an elliptical
trajectory. In this case I find, using the standard equations,
\begin{equation}
v^2=GM\left ({2\over r}-{1\over {\mathcal A}}\right), \qquad
{1\over 2}rv^2=A={1\over 2}{\mathcal B}\sqrt {{GM}\over {\mathcal A}},
\label{6.12}
\end{equation}
where ${\mathcal A,\; B}$ are the major and minor semi-axes, and the areal
velocity $A$ is a constant by the conservation of angular momentum.  In this
case I find ${\mathcal A}\approx 1.0437$ and ${\mathcal B}\approx 1.0216$.
In as much as the semilatus rectum parameter $p=1$,
the latus rectum itself is clearly defined by the intersection of this ellipse
with the unit circle (Schwarzschild trajectory). The latus rectum is the line
through the focus (origin in this case) which is perpendicular to
the semimajor axis.  The eccentricity is $e\approx 0.20462$.

\acknowledgements
The author is pleased to acknowledge helpful conversations with S. Habib,
P. O. Mazur, E, Motolla, and M. M. Nieto.

\end{multicols}


\begin{references}
\bibitem{Peb} P. J. E. Peebles, {\it Principles of Physical Cosmology}
(Prince\-ton University Press, Princeton, NJ, 1993).
\bibitem{Br} G. D. Birkhoff, {\it Relativity and Modern Physics} (Harvard
University Press, Cambridge, Mass., 1923)
\bibitem{T} R. C. Tolman, {\it Relativity, Thermodynamics and Cosmology}
(Oxford University Press, London, 1949)
\bibitem{MTW} C. W. Misner, K. S. Thorne, and J. A. Wheeler, {\it Gravitation}
(Freeman \& Co., San Francisco, 1973).
\bibitem{ADM} R. Arnowitt, S. Deser, and C. W. Misner, L. Witten, ed.
{\it Gravitation, an Introduction to Current Research} (Wiley, New York
1962) pg. 227.
\bibitem{DO} I wish to thank C. C. Dyer and C. Oliwa, astro-ph/0004090,
for drawing this feature to my attention.
\bibitem{Stef} H. Stephani, {\it General Relativity}, J. Stewart, ed.,
(trans.~ of {\it Allgemeine Relativit\"atstheorie} by
M. Pollock and J. Stewart) (Cambridge Univ. Press, 1990, Cambridge).
\bibitem{McV} G. C. McVittie, M.N.R.A.S. {\bf 93}, 325 (1933).
\bibitem{NP} P. D. Noerdlinger and V. Petrosian, Ap.~J. {\bf 168}, 1 (1971).
\bibitem{Pie} B. O. Pierce, {\it A Short Table of Integrals}, no. 222,
(Ginn \& Co., Boston, 1910).
\bibitem{GR} I. S Gradshteyn and I.M. Ryzhik (translated by A. Jeffrey),
{\it Table of Integrals, Series, and Products}, No. 3.167.22, (Academic Press,
New York, 1980), and the 1998 CDROM version.
\end{references}
\end{document}